# Quantitative Interpretations of Energetic Features and Key Residues at SARS Coronavirus Spike Receptor-Binding Domain and ACE2 Receptor Interface

Yanmei Yang[1], Yunju Zhang[2], Yuanyuan Qu[2], Xuewei Liu[3], Mingwen Zhao[2], Yuguang Mu[4*] and Weifeng Li[2*]

1. *College of Chemistry, Chemical Engineering and Materials Science, Collaborative Innovation Center of Functionalized Probes for Chemical Imaging in Universities of Shandong, Key Laboratory of Molecular and Nano Probes, Ministry of Education, Shandong Normal University, Jinan, 250014, China*
2. *School of Physics, Shandong University, Jinan, Shandong, 250100, China*
3. *Division of Chemistry and Biological Chemistry, School of Physical and Mathematical Sciences, Nanyang Technological University, 21 Nanyang Link, Singapore 637371, Singapore*
4. *School of Biological Sciences, Nanyang Technological University, Singapore, 637650*
* Corresponding authors. E-mail: lwf@sdu.edu.cn (W. Li) and ygmu@ntu.edu.sg (Y. Mu)

**Abstract**

The wide spread of coronavirus disease 2019 (COVID-19) has declared a global health emergency. As one of the most important targets for antibody and drug developments, Spike RBD-ACE2 interface has received extensive attention. Here, using molecular dynamics simulations, we explicitly evaluated the binding energetic features of the RBD-ACE2 complex of both SARS-CoV and SARS-CoV-2 to find the key residues. Although the overall ACE2-binding mode of the SARS-CoV-2 RBD is nearly identical to that of the SARS-CoV RBD, the difference in binding affinity is as large as -16.35 kcal/mol. Energy decomposition analyses identified three binding patches in the SARS-CoV-2 RBD and eleven key residues (Phe486, Tyr505, Asn501, Tyr489, Gln493, Leu455 and etc) which are believed to be the main targets for drug development. The dominating forces are from van der Waals attractions and dehydration of these residues. It is also worth mention that we found seven mutational sites (Lys417, Leu455, Ala475, Gly476, Glu484, Gln498 and Val503) on SARS-CoV-2 which unexpectedly weakened the RBD-ACE2 binding. Very interestingly, the most repulsive residue at the RBD-ACE2 interface (E484), is found to be mutated in the latest UK variant, B1.1.7, cause complete virus neutralization escapes from highly neutralizing COVID-19 convalescent plasma. Our present results indicate that at least from the energetic point of view such E484 mutation may have beneficial effects on ACE2 binding. The present study provides a systematical understanding, from the energetic point of view, of the binding features of SARS-CoV-2 RBD with ACE2 acceptor. We hope that the present findings of three binding patches, key attracting residues and unexpected mutational sites can provide insights to the design of SARS-CoV-2 drugs and identification of cross-active antibodies.

# 1. Introduction

To date, the coronavirus disease 2019 (COVID-19) becomes a pandemic threat that has declared a global health emergency of international major concern[1-3]. Since the initial outbreak in China in December 2019, the COVID-19 pandemic quickly spreads nationwide and to more than 80 countries around the world. Patients infected by SARS-CoV-2 showed a range of symptoms that are similar to individuals who were infected by SARS-CoV in 2003 and MERS-CoV in 2012, including fever, dry cough, headache, dyspnea and pneumonia.[4-6] As of 8 March, 2021, more than 116 million cases have been confirmed with the infection worldwide and over 2.5 million cases infected patients have died[7]. Owing to the rapid and global dissemination of COVID-19, the developments of efficient antibodies and drugs are inevitably encountered by worldwide researchers as well as policy makers.

SARS-CoV-2 infection happens by spike glycoprotein (S-protein) receptor binding domain (RBD) on the virus envelope recognizing and binding to the human cellular receptor angiotensin-converting enzyme 2 (ACE2). Recent studies highlighted the important role of ACE2 in mediating the entry of SARS-CoV-2.[8-11] Experimental evidence reveals that only Hela cells that expressing ACE2 are susceptible to SARS-CoV-2 whereas those without ACE2 expression are not. Previous theoretical studies of the complex of the SARS-CoV-2 RBD and ACE2 have identified some residues that are potentially involved in the interaction.[12] However, a quantitative analysis of the actual residues that mediate the interaction remained unclear. More importantly, despite the high similarity between SARS-CoV-2 and SARS-CoV, isolated SARS-CoV monoclonal antibodies are reported to be unable to neutralize SARS-CoV-2.[13, 14] Therefore, a deep understanding of the binding energetic characteristics between SARS-CoV-2 RBD and ACE2 receptor is critical for antibody and drug developments. Especially that, the key residues that mediate the interactions between RBD and ACE2 are, principally, the targets for therapeutic interventions.

This situation highlights the urgent necessity and feasibility of establishing a quantitative model that reveals structure differences between the SARS-CoV and SARS-CoV-2 RBD complexing with the receptor and explore how the intrinsic sequence and structure differences between the SARS-CoV and SARS-CoV-2 RBDs regulate the binding with ACE2. The cryo-electron microscopy structures of the SARS-CoV and SARS-CoV-2 S-protein binding to ACE2 have recently been reported by several independent groups.[8, 9, 14] These structures enable us to theoretically interpret the key residues dominating the total binding affinity (ΔG) at a higher precision which is critical information for medical countermeasure development, especially antibody and drug developments.

## 2. Model Construction and Computing Details

To quantitatively elucidate the interactions between the SARS coronavirus RBD and ACE2, we performed molecular dynamics (MD) modelling of the binding between the ACE2 and RBDs of SARS-CoV and SARS-CoV-2. The coordinates (as depicted in Fig. 1a and 1b) for the SARS-CoV-2 RBD-ACE2 complex were adopted from crystal structure by Zhou et. al., with PDB code 6M17.[15] While for SARS-CoV RBD-ACE2, the PDB code is 2AJF.[16] Without losing the validity, the full length was truncated to contain the peptidase domain (PD) of ACE2 and RBDs of SARS-CoV/SARS-CoV-2. The complex was solvated in a cubic water box with a dimension of 14.3×14.3×14.3 nm$^3$. Sodium ions were added to neutralize the net charge of proteins, resulting in totally 300,631 and 288,426 atoms for SARS-CoV and SARS-CoV-2 models.

All MD simulations were conducted with the GROMACS package.[17] The AMBER03 force field[18] was employed for proteins and salt ions. The TIP3P water model[19] was utilized in all the simulated systems. The electrostatic interactions were treated using the particle mesh Ewald (PME) method.[20, 21] All the bonds involving hydrogen atoms were maintained at their constant equilibrium lengths with the LINCS algorithm.[22] Each system was first energy-minimized with the steepest descent algorithm, followed by 10 ns of pre-equilibration with position restraints applied on the Spike RBD and PD atoms conducted in the NPT ensemble (1atm, 300 K) using velocity-rescale thermostat,[23] where random initial velocities were assigned to protein and solvent atoms. For each system, five parallel trajectories of 1 μs were conducted for data-collection. A time step of 2.0 fs was used, and data were collected every 1 ps.

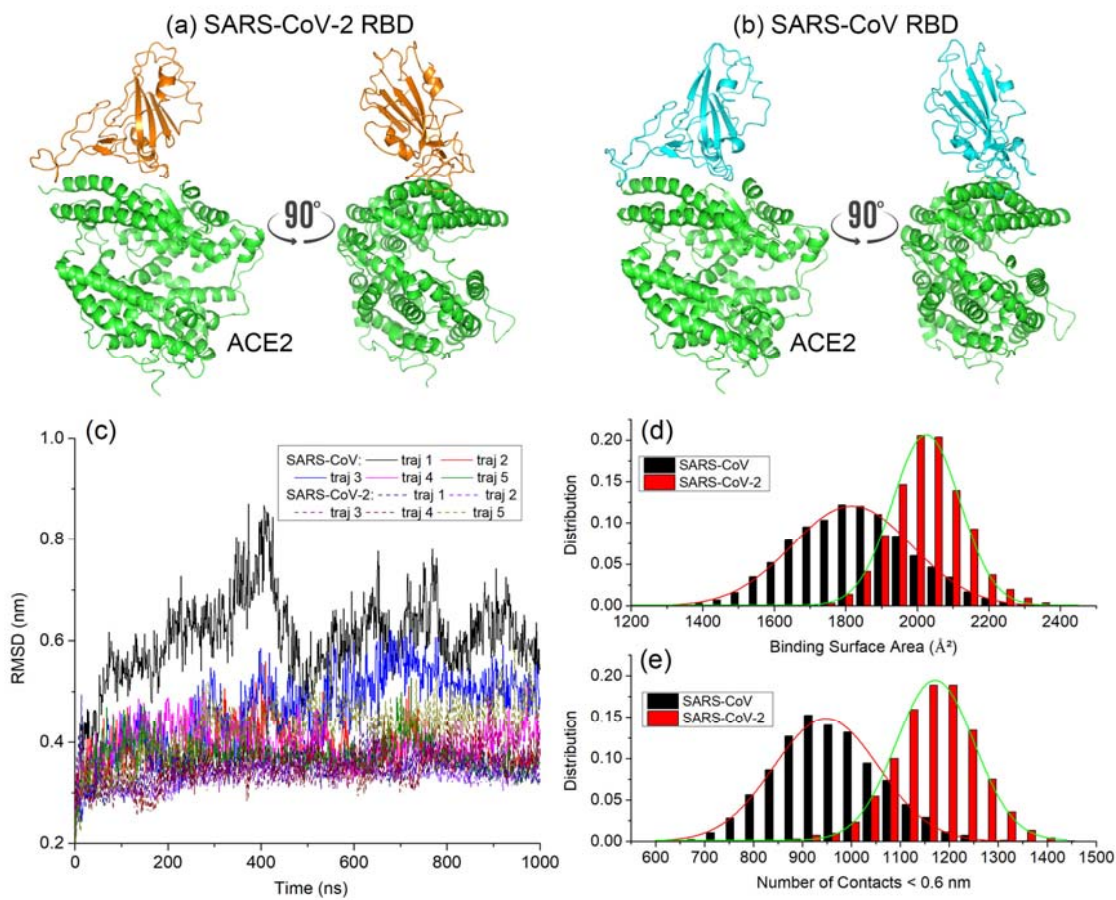

**Figure 1**. Overall model illustrations of ACE2 bound to (a) SARS-CoV-2 RBD and (b) SARS-CoV RBD. The ACE2 is shown in green, SARS-CoV-2 RBD in orange and SARS-CoV in cyan. (c) Time evolutions of the root mean squired displacement (RMSD) of RBD-PD heavy atoms with respect to the Cryo-EM structure (for SARS-CoV-2) and crystal structure (for SARS-CoV), respectively. (d) Distributions of the binding surface area and (e) number of contacts of heavy atoms (0.6 nm cutoff). In (d) and (e), The solid lines represent single peak fitting by Gaussian function.

## 3. Results and Discussions

### 3.1 Structural Stability of Spike RBD-ACE2 Binding Complex

To check the structural stability of RBD-ACE2 binding complex, we firstly calculated the root mean squared displacement (RMSD) of heavy atoms in the two models with respect to the crystal structures. As depicted in Fig. 1c, the SARS-CoV-2 RBD-ACE2 complex demonstrated a considerably high stability where the RMSD values are usually less than 0.5 nm in five trajectories. This phenomenon reflects that SARS-CoV-2 binding with ACE2 is considerably stable which effectively restraints the protein chain diffusion. The RBD of SARS-CoV-2 contains several loop regions at the receptor-binding motif (RBM), which functions in the COVID-19 infection are not well known. In the present simulations, the structures of these loops are found to be well conserved. In contrast, the RMSD for SARS-CoV RBD-ACE2 is slightly larger except trajectory 1. For trajectory 1, the maximum value of RMSD reached 0.85 nm at 400 ns during the simulations (we have to mention that the larger structural fluctuations indeed do not weaken the RBD-ACE2 binding affinity which is discussed in the following). Combining these, we hypothesize that SARS-CoV-2 RBD binds to ACE2 in a stronger manner than SARS-CoV because the interactions effectively restraint the loop mobility.

**3.2 Interaction Features of Spike RBD-ACE2 Interface**

From the above analyses, the complex structures are considerably stable in the ten 1 μs trajectories. To find the quantitative differences at the RBD-PD interface between SARS-CoV and SARS-CoV-2, we calculated the binding surface area (BSA) between RBD and PD based on these sampled structures. For SARS-CoV-2 as shown in Fig. 1d, the trajectories reveal a large binding interface with a BSA of 2,028 Å$^2$ (1,018 Å$^2$ on the RBD and 1,010 Å$^2$ on the PD). While for SARS-CoV, a clearly smaller BSA of 1,822 Å$^2$ (931 Å$^2$ on the RBD and 891 Å$^2$ on the PD) is observed. Generally, we expect that SARS-CoV-2 RBD-ACE2 binding involves more residues from the two counterparts thus the binding is more intimate.

Consistent with the larger BSA of ACE2 with SARS-CoV-2, the two counterparts also formed more contacts (here a contact is recognized when a heavy atom from ACE2 is within 0.6 nm to a heavy atom of RBD). The number of contacts ($N_c$) were calculated and depicted in Fig. 1e. It is found that there are averagely 1166 contacts between SARS-CoV-2 and ACE2. While for SARS-CoV, the corresponding value is only 955, resulting in difference ratio of 22.1%. In addition, for both the BSA and $N_c$, the values disperse in a wide range for SARS-CoV RBD-ACE2 complex. While for SARS-CoV-2 RBD-ACE2, the values are more localized. This difference reveals a fact that SARS-CoV-2 complexing with ACE2 in a more stable mode than SARS-CoV.

**Table 1**. Spike RBD-ACE2 binding affinity (unit: kcal/mol) of SARS-CoV and SARS-CoV-2 in ten trajectories. $E_{vdW}$ and $E_{ele}$ represent direct van der Waals and electronic interactions between Spike RBD and ACE2. $G_{surf}$ represents the non-polar solvation energy calculated from solvent-accessible surface area model and $G_{GB}$ is the polar solvation energy obtained from Generalized Born model. The corresponding values in the brackets are standard deviations.

| Trajectory | | Total ΔG | $E_{vdW}$ | $G_{surf}$ | $E_{ele}$ | $G_{GB}$ |
|---|---|---|---|---|---|---|
| SARS-CoV | 1 | -33.21 (6.85) | -78.54 (7.75) | -10.65 (1.04) | -674.25 (46.46) | 730.23 (46.04) |
| | 2 | -37.36 (8.21) | -87.16 (8.88) | -12.08 (1.10) | -623.47 (44.56) | 685.35 (44.85) |
| | 3 | -39.90 (7.05) | -87.17 (7.70) | -11.86 (1.10) | -616.78 (40.28) | 675.92 (40.47) |
| | 4 | -35.45 (6.45) | -84.20 (8.12) | -11.85 (1.01) | -634.01 (41.49) | 694.62 (40.21) |
| | 5 | -31.95 (6.60) | -81.58 (7.39) | -10.67 (0.95) | -634.37 (38.71) | 694.68 (39.15) |
| SARS-CoV-2 | 1 | -48.39 (6.97) | -108.61 (7.58) | -14.27 (0.90) | -704.02 (63.18) | 778.50 (63.82) |
| | 2 | -59.41 (8.79) | -107.91 (8.10) | -14.94 (0.97) | -703.64 (43.60) | 767.08 (42.62) |
| | 3 | -50.68 (8.27) | -100.67 (7.97) | -13.52 (0.96) | -666.29 (43.48) | 729.81 (41.37) |
| | 4 | -49.38 (7.86) | -104.50 (6.39) | -13.83 (0.75) | -610.82 (45.43) | 679.77 (44.11) |
| | 5 | -51.75 (6.53) | -97.55 (4.92) | -13.31 (0.62) | -688.01 (49.08) | 747.12 (47.67) |

The massive simulated trajectories enable us to evaluate the binding affinity (ΔG) between Spike RBD and ACE2. Consistent with experimental results,[14, 24-26] the theoretical predicted ΔG reveals that SARS-CoV-2 RBD binds ACE2 with higher affinity than SARS-CoV-1 (Table 1). For SARS-CoV-2, the average ΔG of RBD-ACE2 are calculated to be -51.92±7.68 kcal/mol from five trajectories. This value is in line with recently theoretical predictions with PBSA method.[27] While for SARS-CoV, the values are -35.57±7.03 kcal/mol, respectively. Thus, the complexation of ACE2 with SARS-CoV-2 RBD is considerably stronger than the SARS-CoV which is well consistent with the experimental findings[14, 24-26]. We also noticed that the enhanced binding was physically caused by the vdW attractions ($E_{vdW}$) and dehydration of unpolar residues ($G_{surf}$) which dominantly contributed to the total ΔG. In contrast, although the electronic attractions ($E_{ele}$) between Spike RBD and ACE2 are considerably strong, the complexation process is rather energy-consumable because the energy release ($E_{ele}$) cannot compensate the dehydrations of polar and charged residues ($G_{GB}$). This is also in line with experimental observations by Bloom and co-workers that reducing the polar character of interfacial residues often enhance affinity[24]. From this result, we expect that more hydrophobic or aromatic residues are located at the RBM of SARS-CoV-2 Spike than the SARS-CoV.

**3.3 Energy Decomposition Analyses of Interfacial Residues**

As stated, amongst all the residues at RBD-ACE2 interface, the identification of key residues that drive the binding process is essential for drug development because not all of them dominate the binding. Thus, we calculated the contributions of each residue of Spike RBD and ACE2-PD domains to total $\Delta G$ (denoted as $\Delta G^{res}$). Fig. 2 depicts the contributions of key residues of Spike RBD to total $\Delta G$ which are expressed in increasing numerical order of $\Delta G^{res}$. Here, key residue is recognized when $|\Delta G^{res}| > 0.3$ kcal/mol. Negative $\Delta G^{res}$ stands for an attractive residue, and a positive $\Delta G$ indicates a repulsive residue.

Totally there are six residues of SARS-CoV-2 contributing more than -2 kcal/mol to total $\Delta G$ which involve: Phe486(-5.15), Tyr505(-4.03), Asn501(-3.37), Tyr489(-2.66), Gln493(-2.53) and Leu455(-2.29). Here the values in the bracket indicate the corresponding $\Delta G^{res}$ in kcal/mol. For SARS-CoV, there are less attracting residues. Moreover, the first two key residues, Thr487 and Pro462, contribute only -3.51 and -3.47 kcal/mol to total $\Delta G$, which is clearly weaker than their counterparts (Phe486 and Tyr505) in SARS-CoV-2. In contrast to these attracting residues, there still exist numerous residues with positive $\Delta G^{res}>0.3$ mcal/mol, acting like "hindrance residues" to block the RBD-ACE2 binding. For SARS-CoV-2, these include Glu484(0.97), Glu406(0.61), Asp405(0.52), Lys417(0.49) and Gln506(0.36). For SARS-CoV, there is one more which involve Ser432(0.73), Lys390(0.71), Lys465(0.60), Asp392(0.54), Asp480(0.44) and Asp393(0.35). Although these residues act preventing the attraction between RBD and ACE2, the capability is low because the corresponding $\Delta G^{res}$ values are all below 1 kcal/mol.

Very recently, several experimental groups independently reported the evolution studies of SARS-CoV-2 in the immune population.[28-30] They found that SARS-CoV-2 with several substitutions (K417, E484 and N501) escapes neutralization from highly neutralizing COVID-19 convalescent plasma, amongst which E484 accounted for much of the effect.[28-31] Our present findings of the highest $\Delta G^{res}$ value of E484, indicate that at least from the energetic point of view, the mutation of E484, has beneficial effects on ACE2 binding.

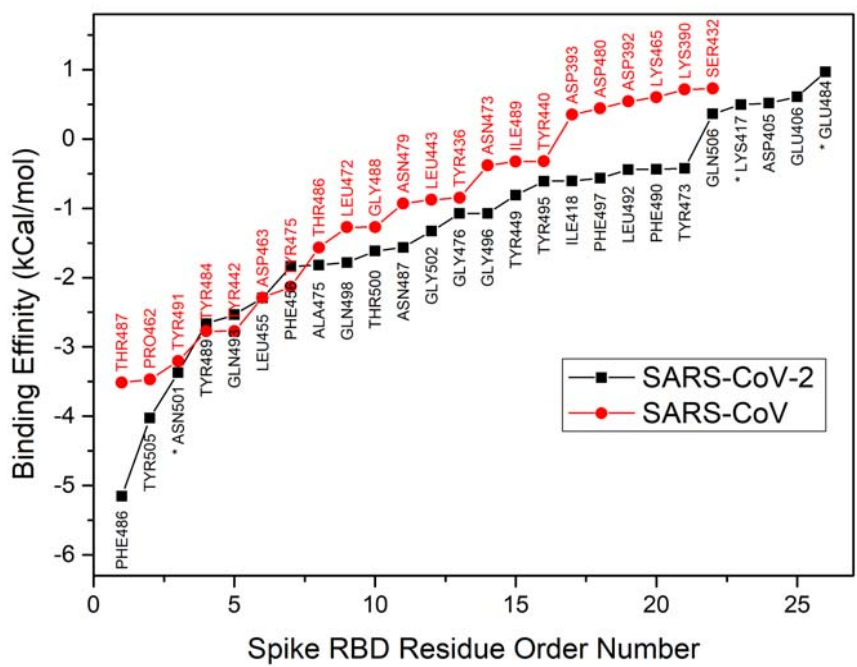

**Figure 2**. Key residues at the RBD-ACE2 complex binding interface of SARS-CoV-2 and SARS RBD. Here, decomposed binding affinity with $|\Delta G| > 0.3$ kcal/mol is recognized as key residues. The stars indicate three mutation sites, K417, E484 and N501 found in SARS-CoV-2 lineage in South Africa (Nextstrain clade 20H/501Y.V2).[32]

**3.4 Key Residues Distribute in Three Spatially Separated Patches**

According to the spatial distributions, these residues are classified in to three patches as shown in Fig. 3:

**Patch 1**. Four residues, including Ala475, Phe486, Asn487 and Tyr489, form a dominating binding patch (Patch-1 as indicated in Fig. 3a). In the RBD, a total of nine cysteine residues exists, eight of which form four disulfide bonds. Among these four pairs, three are inside RBD core (Cys336–Cys361, Cys379–Cys432 and Cys391–Cys525). Other than these, one remaining pair (Cys480-Cys488) connects the loops in the distal end of the RBM. Because the TOP1 residue Phe486 just locates in the distal end, this Cys480-Cys488 disulfide bond is believed to be important for the Phe486 to bind with ACE2 because it restrains the conformation of the loop. Quantitatively, summarizing the decomposed $\Delta G^{res}$ of the patch 1 residues, they contribute 21.6% to the total $\Delta G$.

For SARS-CoV-2, the Phe486 has 7 counterparts on ACE2, in detail, Thr20, Gln24, Phe28, Leu79, Met82, Tyr83, Pro84. The Phe486 site of SARS-CoV-2 corresponds to Leu472 of SARS-CoV. Due to the relatively smaller size, the counterpart residues only involve Glu75, Thr78, Leu79, Met82 and Tyr83. Because both Phe486 and Leu472 are hydrophobic residues, the driving force mainly comes from van der Waals attractions and de-hydration process. In addition, the Phe486 is capable to form π-π stacking with aromatic residues like Phe28, Tyr83 and Pro84 which is commonly believed to be strong interactions.

**Patch 2**. Besides Phe486, the Tyr505 (TOP2) and Asn501 (TOP3) also contribute -4.03 and -3.37 kcal/mol to the total binding. Together with Gln498 and Thr500, they form the second binding patch (Patch-2 in Fig. 3a). These two residues correspond to Tyr491 (TOP3) and Thr487 (TOP1) respectively in SARS-CoV with $\Delta G^{res}$ of -3.20 and -3.51 kcal/mol. Generally, the locations of Tyr505+Asn501 and Tyr491+Thr487 coincide at the RBM surface. Quantitatively, the patch 2 contributes 20.8% to the total $\Delta G$.**Patch 3**. At the middle of the RBM surface, the third binding patch (Patch-3 in Fig. 3a) is recognized which is composed of Leu455(-2.29), Phe456(-1.84) and Gln493(-2.53). While for SARS-CoV, this patch only contains Tyr442 with a $\Delta G^{res}$ of -2.77 kcal/mol. The patch 3 contributes 12.8% to the total $\Delta G$.

The above three patches play a major role in the complexing of RBD-ACE2. Other than these residues, there are more residues in SARS-CoV-2 participating in the binding with ACE2 compared to SARS-CoV. In detail, there are 21 residues of SARS-CoV-2 contributing more than -0.3 kcal/mol to the total $\Delta G$, compared to only 16 residues in SARS-CoV.

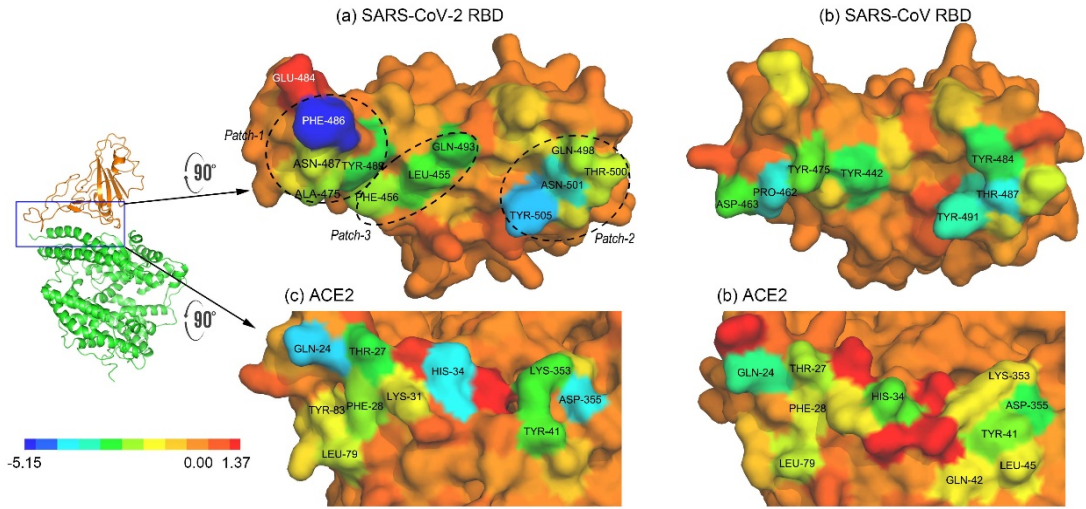

**Figure 3**. Spatial distributions of the key residues at the Spike RBD and ACE2-PD interface. The surface of the residues is colored according to their decomposed energetic contributions ($\Delta G^{res}$) to the binding affinity (blue to red: attractive residues to repulsive residues).

It is worth mention that, Bloom and co-workers have reported that mutations of Gln493, Gln498 and Asn501 can enhance ACE2 affinity[24]. These three residues have also been observed in the present studies where Gln494 belongs to patch-3, Gln498 and Asn501 belong to patch-2.

As the counterparts of Spike RBD, the key residues in ACE2-PD are mostly similar when binding to SARS-CoV-2(Fig. 3c) and SARS-CoV (Fig. 3d). For instance, the TOP3 key residues uniformly involve Gln24, His34 and Asp355 in the two cases. However, we noticed that most of these counterpart residues bind in a stronger manner with SARS-CoV-2 RBD than the SARS-CoV. Especially for the TOP3 residues, the binding energy difference is as large as about –1 kcal/mol for each residue, indicating more intimate binding between SARS-CoV-2 and ACE2.

**3.5 Sequence Alignment According to Decomposition Binding Energy**

The whole sequence of SARS-CoV-2 shares around 80% similarities to SARS-CoV. At the RBD domain, there are 35% mutations (41 out of 116). However, the specific roles of the Spike RBD, especially for the mutation sites, are yet to be well documented. Summarizing all the present data, we classified the Spike RBD sequence into three types according to their specific roles in the Spike RBD-ACE2 binding (Fig. 4).

```
                           ▼▲
              ▲    ○○      ○○
SARS-CoV-2 RBD 400 FV I R G D E V R Q I A P G Q T G K I A D Y N Y K L P D D F T G C V I A W N S N 439
SARS-CoV   RBD 387 FV V K G D D V R Q I A P G Q T G V I A D Y N Y K L P D D F M G C V L A W N T R 426
                      ○  ○○

                     ▲           ○     ▼▼▲                         ▲ ▼▼   ▲
                                 ○     ○○                        ○ ○○   ○
SARS-CoV-2 RBD 440 N L D S K V G G N Y N Y L Y R L F R K S N L K P F E R D I S T E I Y Q A G S T P 479
SARS-CoV   RBD 427 N I D A T S T G N Y N Y K Y R Y L R H G K L R P F E R D I S N V P F S P D G K P 466
                       ○         ○          ○○                                        ○○○

                         ▼ ▲▲ ▲▲   ▲▲ ▲▲▲▲▼      ▲ ▼
                         ○ ○○ ○○ ○ ○○○○ ○○○○    ○○
SARS-CoV-2 RBD 480 C N G V E G F N C Y F P L Q S Y G F Q P T N G V G Y Q P Y R V V V L S F E 516
SARS-CoV   RBD 467 C T P - P A L N C Y W P L N D Y G F Y T T T G I G Y Q P Y R V V V L S F E 502
                          ○        ○○       ○○   ○○○○  ○
```

**Figure 4**. Sequence alignment of SRAS-CoV-2 and SARS-CoV RBD. Circles indicate key residues mediating the binding (|ΔG$^{res}$|>0.3 kcal/mol). For residues in SARS-CoV-2, those enhancing the binding affinity (ΔΔG$^{res}$<-0.3 kcal/mol) than SARS-CoV are indicated by up-triangle, while those weakening the binding ((ΔΔG$^{res}$>0.3 kcal/mol)) are indicated by down-triangle.

**Type 1**. Strengthen attraction (residues with decreased negative ΔG$^{res}$): this represents residues in SARS-CoV-2 RBD forming stronger attraction with ACE2 than SARS-CoV, which involves Ile418, Phe456, Tyr473, Phe486, Asn487, Tyr489, Phe490, Leu492, Gln493, Tyr495, Gly496, Phe497, Tyr505;

**Type 2**. Weaken repulsion (residues with decreased positive ΔG$^{res}$): this represents residues in SARS-CoV-2 RBD weakening the repulsion to ACE2 than SARS-CoV, which involves Arg403, Val445, Thr478, Ser494;

**Type 3**. Hinder binding (residues with increased ΔG$^{res}$): this represents residues in SARS-CoV-2 playing a negative role to the RBD-ACE2 binding, which involves Lys417, Thr453, Leu455, Ala475, Gly476, Glu484, Gln498, Val503, Gln506.

The type 1 and type 2 residues are the essential residues for the increased binding strength of SARS-CoV-2 than SARS-CoV with ACE2. While for type 3 residues, it is unexpected that seven of them (except Tyr453 and Gln506) are mutation sites but reversely hinder the Spike RBD-ACE2 binding. This is unexpected because whether these mutational sites have functional relevance are yet to be known.

## 4. Conclusions

To summarize, we have systematically simulated the binding structures of RBD-ACE2 of both SARS-CoV and SARS-CoV-2. By Generalized Born model, the key residues that dominating the complexation process have been found and discussed. At the RBD-ACE2 interface, three binding patches in the SARS-CoV-2 RBD and eleven key residues (Phe486, Tyr505, Asn501, Tyr489, Gln493, Leu455 and etc.) have been identified which are believed to be the main targets for drug development. Seven mutational sites (Lys417, Leu455, Ala475, Gly476, Glu484, Gln498 and Val503) on SARS-CoV-2 unexpectedly weakened the RBD-ACE2 binding. Interestingly the most repulsive residue at the RBD-ACE2 interface, Glu484, is found to be mutated in the latest UK variant, B1.1.7[31] to cause virus neutralization escapes from highly neutralizing COVID-19 convalescent plasma. Further efforts are highly required to probe whether these sites have other functional relevance. The present findings gave a deep understanding of the binding features between RBD and ACE2, providing important structural insights to the design of SARS-CoV-2 drugs and identification of cross-active antibodies.

**Acknowledgements**

This work is supported by the Natural Science Foundation of Shandong Province (ZR2020MB074, ZR2020JQ04) and National Natural Science Foundation of China (11874238).